\documentclass[showpacs,twocolumn,prl,floatfix,nobalancelastpage]{revtex4}
\usepackage{graphicx}
\usepackage{times}
\usepackage{bm}
\usepackage{amsmath}

\begin{document}

\date{\today}
\title{Husimi functions at dielectric interfaces:
Inside-outside duality for optical systems and beyond}

\author{Martina Hentschel}
\author{Henning Schomerus}
\affiliation{
Max-Planck-Institut f\"{u}r Physik komplexer Systeme,
N\"othnitzer Str. 38, D-01187~Dresden, Germany}
\author{Roman Schubert}
\affiliation{
Service de Physique Th\'eorique, CEA/DSM/SPhT, Unit\'e de recherche associ\'ee au CNRS,
CEA/Saclay,
F-91191 Gif-sur-Yvette Cedex, France}

\begin{abstract}
We introduce generalized Husimi functions at the interfaces of dielectric systems.
Four different functions can be defined, corresponding to the incident and
departing wave on both sides of the interface. These functions allow to identify
mechanisms of wave confinement and escape directions in optical
microresonators, and give insight into the structure of resonance
wave functions.
Off resonance, where systematic interference can be neglected, the
Husimi functions are related by Snell's law and Fresnel's coefficients.
\end{abstract}

\maketitle

Optical microresonators receive growing interest over the last years,
because of the
intricate interplay of shape (leading to irregular
classical ray dynamics), openness
of the system (offering means of excitation and escape),
and the wave nature of the field.
This interplay, together with the promising prospect of
applications in future communication devices, has stimulated
experiments \cite{franzosen,sangwook,rexstone}
as well as theoretical investigations \cite{noeckelstone,annbillpre}
that were based on concepts well-known from
scattering theory, classical ray dynamics, semiclassics, and quantum chaos
\cite{stoecki}.
A particularly useful tool to study waves in dynamical systems is the Husimi representation
of the wave function in classical phase space.
So far, the Husimi representation was mostly used to study the closed analogues
of optical microsystems, with the dielectric interface being replaced by hard
walls, and the principal confinement and radiation directions have been
inferred by adding the laws of reflection and refraction by hand.
Most notably, efforts in this direction suffer from the fact that the
incident and emerging wave components cannot be discriminated
by the conventional Husimi representation.
The reasons for these fundamental shortcomings
arise from the facts that the wave function of the dielectric system is
only partially confined by the
internal reflection at the refractive index boundary,
and that it is affected by the different nature of the boundary conditions, which are neither of
Dirichlet nor of von-Neumann type but of a mixed type that follows
from Maxwell's equations (both the wave
function
and its derivative are non-vanishing at the interface).

In this paper we introduce four Husimi
functions appropriate for dielectric interfaces, corresponding to the intensity
of incident and emerging waves at both sides of the
interface.
In the regime of ray optics it will be demonstrated that these Husimi functions
are related across the interface via Fresnel's formulas, with phase space being deformed
according to Snell's law.
This connection can be seen as a new variant of
the inside-outside duality \cite{IOS}.
However,
ray optics only applies when the wave length is short
and when systematic interference effects can be neglected.
The Husimi functions do not require these limitations
and develop their full predictive power
especially when systematic interference effects lead to strong
deviations from Fresnel's and Snell's laws.
In particular, ray optics breaks down close to resonances, where
the internal part of the scattering wave function is
known to be almost
independent of the incoming wave that excites the system.
The Husimi functions still provide an accurate representation of the wave
function (in particular,
they nicely display the radiation directions of the field).
We illustrate these features using the dielectric circular
disk and an annular system as
examples \cite{remark_pol}.

\begin{figure}
\includegraphics[width=0.35\textwidth]{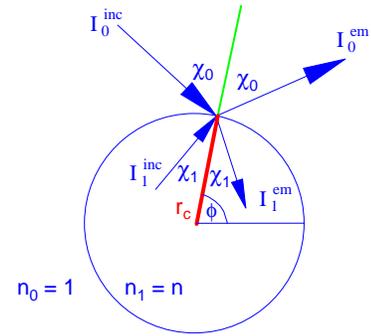}
  \caption{Refractive-index boundary of a dielectric system.}
  \label{fig_boundary}
\end{figure}

{\em Microresonators and ray optics.}
Consider the light that illuminates and  permeates
a dielectric system confined by a refractive-index boundary, as shown in
Fig.~\ref{fig_boundary}.
The disk (and also the annulus) are confined
by  a circular dielectric interface of radius $r_c\equiv 1$.
Polar coordinates $r$, $\Phi$ will be used to parameterize position space.
We distinguish four wave components: The incident (inc) wave and the emerging
(em) wave on both sides of the interface (region 0 with refractive index $n_0=1$
outside the microresonator,
region 1 with $n_1\equiv n$ inside the microresonator).
In the regime of ray optics, the wave is represented by rays,
where the angles of
incident and
emerging rays are related by the laws of reflection and Snell's law,
$n\sin\chi_1= \sin \chi_0$.
For our circular interfaces Snell's law is equivalent to conservation of the
angular-momentum variable $m=k_j\sin\chi_j$,  where $k_0$ and $k_1=n
k_0$ are the wave number in each region.
The ray
intensities on either side follow from Fresnel's laws,
\begin{eqnarray}
I^{\rm em}_{0} & = & R_0 (\chi_0) \: I^{\rm inc}_{0} + T_1 (\chi_1) \: I^{\rm inc}_{1}
,
\label{intausaus}\\
I^{\rm em}_{1} & = & R_1 (\chi_1) \: I^{\rm inc}_{1} + T_0 (\chi_0) \: I^{\rm inc}_{0}
.
\label{intinin}
\end{eqnarray}
The reflection and transmission coefficients $R_i$ and $T_i$ are related
by
\begin{equation}
R_0(\chi_0) = R_1 (\chi_1) \equiv R \:,
\quad T_0(\chi_0) = T_1 (\chi_1) = 1-R \:,
\label{rtausseninnen}
\end{equation}
with \cite{remark_pol}
$R= \sin^2(\chi_1-\chi_0) / \sin^2(\chi_1+\chi_0)$.

{\em Husimi functions at a dielectric interface.}
The Husimi function at the system boundary of closed
systems was introduced in
Refs.\ \cite{stattSchubertBaecker,Schubert}
by projection of the conventional Husimi function from full phase space
[coordinates $(r, \Phi)$, momentum $(k_j\sin\chi_j, k_j\cos\chi_j)$]
onto the reduced phase space
at the boundary $r=r_c$
with coordinates $\phi=\Phi$ and
$\sin \chi_j$ \cite{remark_nonspherical}.
The four different
Husimi functions (corresponding to the incident and emerging
wave at both sides of a dielectric interface)
can be constructed by the same procedure
when the appropriate boundary conditions are employed.
The intensities
$I(\phi, \sin \chi) = H(\phi, \sin \chi) \,  d \phi  \, d\sin \chi$
will turn out to be related by the laws (\ref{intausaus}), (\ref{intinin})
when ray optics applies, but also accurately describe the wave function
when ray-optical relations across the interface break down, as expected, e.g.,
for resonances.

The conventional Husimi function for a given
wave function  $\Psi(r, \Phi)$ of the dielectric system
is obtained as the overlap
with a wave packet with minimal uncertainty
in the variables
$(r, \Phi)$ for real space and $(k_j\sin \chi_j,k_j\cos \chi_j)$
for momentum space.
The projection
onto the
boundary can be formulated rigorously \cite{Schubert}: 
The wave function is expressed by means of advanced and retarded
Green's functions, which in turn allow to distinguish between incident and
emerging waves.
Green's formula is then used to express the
solution $\Psi_j$ of the Helmholtz equation in
region $j=0$, $1$ as an integral over the
boundary, involving both $\Psi_j$ and its normal (radial) derivative $\Psi'_j$.
A semiclassical (saddle-point) approximation then allows to
identify in these expressions the following four different
Husimi functions on the interface,
\begin{eqnarray}
&&H^{\rm inc(em)}_{j}(\phi,\sin\chi) \nonumber \\
&&= \frac{k_j}{2 \pi}\left|
                (-1)^j {\cal F}_jh_j(\phi,\sin\chi)
                     + (-)\frac{i}{k_j{\cal F}_j} h_j'(\phi,\sin\chi)
                     \right|^2
,\quad~
\label{hus}
\end{eqnarray}
with the angular-momentum dependent weighting factor
${\cal F}_j = \sqrt{ n_j\cos \chi_j}$ \cite{remark_cutoff}. Here the functions
\begin{eqnarray}
h_j&=&\int_0^{\infty} r\,dr \int_0^{2\pi} d\Phi\,\Psi_j(r,\Phi)\xi(r,\Phi;\phi,\sin\chi),
\\
h_j'&=&\int_0^{\infty} r\,dr \int_0^{2\pi} d\Phi\,\Psi'_j(r,\Phi)\xi(r,\Phi;\phi,\sin\chi)
\end{eqnarray}
are overlaps with the minimal-uncertainty wave packet
\begin{eqnarray}
\xi(r,\Phi;\phi,\sin\chi) &=&
                \sum_l
       e^{-\frac{1}{2 \sigma} (\Phi + 2 \pi l -\phi)^2 - i
k\sin\chi (\Phi + 2 \pi l)}
\nonumber
\\
&&\times
(\sigma \pi)^{-\frac{1}{4}}  \delta(r-r_c)
\end{eqnarray}
(a periodic function in $\Phi$), which is restricted to the interface
and
centered around
$(\phi, \sin\chi)$.
The parameter
$\sigma$ controls its extension
in $\phi$-direction,
thereby also fixing the uncertainty
in $\sin \chi$.
We set $\sigma = \sqrt{2} / k_1$. The scaling with $k_1$ results
in the same resolution in $\phi$ for all four Husimi functions.

{\em Inside-outside duality.}
As a consequence of the boundary conditions
derived from Maxwell's equations we find the identities
$h_0(\phi,\sin\chi_0)=h_1(\phi,\sin\chi_1)$,
$h_0'(\phi,\sin\chi_0)=h_1'(\phi,\sin\chi_1)$,
where the angles $\chi_i$ are related by Snell's law.  From
these relations it follows that our Husimi functions strictly fulfill
the condition of intensity conservation,
\begin{eqnarray}
&&n H^{\rm em}_0(\phi,\sin\chi_0)
+H^{\rm em}_1(\phi,\sin\chi_1)
\nonumber\\
&=&
n H^{\rm inc}_0(\phi,\sin\chi_0)
+H^{\rm inc}_1(\phi,\sin\chi_1)
\:,
\end{eqnarray}
where the factor $n=d\sin\chi_0/d\sin\chi_1$ accounts for the
phase-space deformation by Snell's law.
Additional relations between the Husimi functions can be anticipated
in the regime of ray optics: The intensities on one side of the interface
should be related to the intensities on the other side by
Eqs.\ (\ref{intausaus}), (\ref{intinin}).
We then expect validity of the resulting inside-outside duality relations
\begin{eqnarray}
H^{\rm em}_{0} &\approx& S(H^{\rm em}_{0})= \frac{1}{n}\frac{1-2R}{1-R} \: H^{\rm inc}_{1} +
\frac{1}{n}\frac{R}{1-R} \: H^{\rm em}_{1}
,\quad
\label{reconstaus}
\\
H^{\rm inc}_{0} &\approx&  S(H^{\rm inc}_{0})=
-\frac{1}{n}\frac{R}{1-R} \: H^{\rm inc}_{1} + \frac{1}{n}\frac{1}{1-R} \: H^{\rm em}_{1}
,\quad
\label{reconstein}
\end{eqnarray}
which express the Husimi functions in  region $0$ by the Husimi
functions in  region $1$. The notation $S(\cdot)$ indicates that the
approximation is of semiclassical (short wave length) nature;
most noticeably, the Husimi functions (intensities) are added incoherently.
The duality relations can also be inverted,
\begin{eqnarray}
H^{\rm em}_{1} &\approx& S(H^{\rm em}_{1})=n \frac{1-2R}{1-R} \: H^{\rm inc}_{0} +
n \frac{R}{1-R} \: H^{\rm em}_{0}
,\quad
\label{reconstaus2}
\\
H^{\rm inc}_{1} &\approx& S(H^{\rm inc}_{1})=
-n\frac{R}{1-R} \: H^{\rm inc}_{0} + n\frac{1}{1-R} \:
H^{\rm em}_{0}
.\quad
\label{reconstein2}
\end{eqnarray}
However, the Husimi functions in region 1 can only be reconstructed
in the strip $|\sin \chi_1| < 1/n$, because
the rest of phase space is isolated from region 0 by total internal
reflection.

The duality relations are exactly fulfilled in two simple cases,
namely, if one incident or emerging wave vanishes
or when the two incident waves have the same intensity
(the two emerging waves then have the same intensity, as well).
We now test the duality relations in more general situations.

\begin{figure}[t]
\includegraphics[width=\columnwidth]{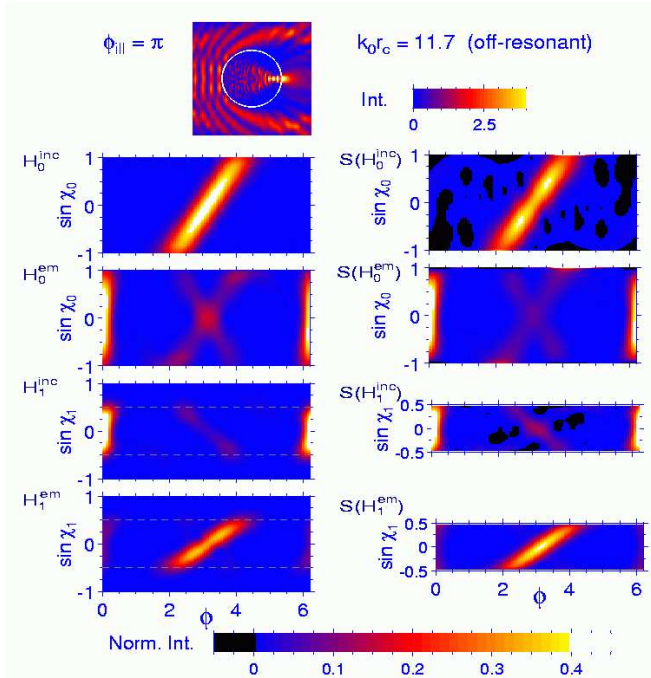}
  \caption{
  Exact (left) and reconstructed (right)
Husimi functions
for the circular dielectric
  disk ($n=2$) illuminated by a plane wave at an off-resonant frequency.
The top panel shows the scattering wave function in real space.
The exact Husimi functions are obtained from Eq.\ (\ref{hus}).
  The reconstructed Husimi functions are obtained
by Eqs.~(\protect\ref{reconstaus})--(\protect\ref{reconstein2}).
  Negative Husimi densities are shown in black.
  The dashed lines in the panels for region $1$
  mark the critical angle of incidence for total internal reflection.
}
  \label{fig_diskoff}
\end{figure}

\begin{figure}[t]
\includegraphics[width=\columnwidth]{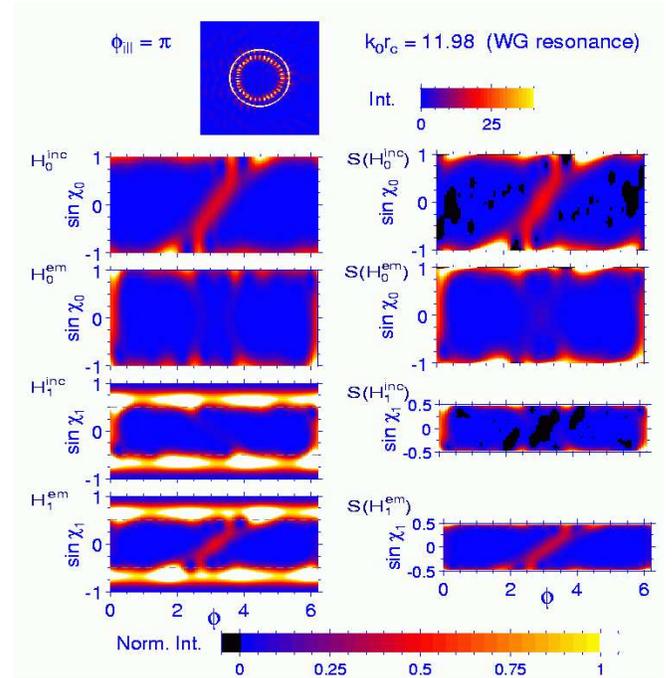}
  \caption{
Same as Fig. \ref{fig_diskoff}, but for illumination at a
  resonance frequency.
}
  \label{fig_diskon}
\end{figure}

The panels on the left in Fig.\ \ref{fig_diskoff}  show the
exact Husimi functions from Eq.\ (\ref{hus}) for the case of the circular disk
which is excited by a plane wave at an off-resonant excitation
frequency.
The illuminating
plane wave is clearly visible  in the
Husimi function $H^{\rm inc}_{0}$, around the polar
angle $\phi=\phi_{\rm ill}=\pi$, while the
focal point
of the dielectric disk
results in a bright spot in $H^{\rm em}_{0}$  that is located around $\phi=0$.
There is a close correspondence between the Husimi functions of the
incident and emerging waves, and the deformation of phase space by the
stretching factor $n$ of Snell's law is clearly visible.

The right panels of  Fig.\ \ref{fig_diskoff}
show for comparison the predictions of Eqs.\
(\ref{reconstaus})--(\ref{reconstein2}).
In the reconstruction we used the slightly modified
semiclassical versions of Fresnel's coefficients
and Snell's laws given in Ref.\ \cite{ghspre}, which are appropriate for the
present case of a curved interface (this results in a slight, but still
noticeable quantitative improvement of the reconstruction).
We observe a good qualitative and quantitative
agreement with the exact Husimi functions.
Regions with unphysical negative intensities
are small.
The most interesting deviations
between the exact and the reconstructed Husimi functions occur
around the central spot at $\phi=\pi, \sin\chi_j=0$,
where the incoherent predictions
of Eqs.\ (\ref{reconstaus})--(\ref{reconstein2})
underestimate the exact Husimi densities $H^{\rm inc}_{0}$, $H^{\rm em}_{0}$, while
they overestimate the intensities $H^{\rm inc}_{1}$, $H^{\rm em}_{1}$
in the same area of phase space.
These deviations
arise from a
Fabry-Perot like systematic interference which is constructive
in backward direction at the presently
chosen frequency. At other frequencies the interference is destructive, and
both cases alternate periodically. 

{\em Resonances.}
Figure \ref{fig_diskon} displays the situation for
excitation at a frequency which is close to a narrow resonance,
a whispering gallery (WG) mode located around $\sin \chi_1 = 0.667$.
The top panel shows that the wave function is now well confined
inside the disk (region 1).
Correspondingly, the Husimi functions $H^{\rm inc}_{1}$ and $H^{\rm em}_{1}$
noticeably exceed the Husimi functions $H^{\rm inc}_{0}$ and
$H^{\rm em}_{0}$.
Moreover, the Husimi functions $H^{\rm inc}_{1}$ and $H^{\rm em}_{1}$ are
dominated by the characteristics
of the resonance wave function and consequentially are
almost independent of the choice of
the exciting wave.
(The remnants of the exciting plane can be identified when comparing
Fig.~\ref{fig_diskon} with Fig.~\ref{fig_diskoff}.)
Hence the reconstructed Husimi functions deviate noticeably from the
exact Husimi functions around $|\sin \chi_1| \sim 1/n$.
This is no surprise since resonances are formed
by systematic constructive interference, and
incoherent ray optics cannot be
expected to apply under these circumstances.
Most importantly, by principle,
the confined wave intensity in the region $|\sin \chi_1| > 1/n$
cannot be reconstructed because classically 
no refracted rays ever reach this region
(which is dark off resonance).
On the other hand, the exact Husimi functions display nicely all the
features of the resonance wave function in phase space.

\begin{figure}[t]
\includegraphics[width=\columnwidth]{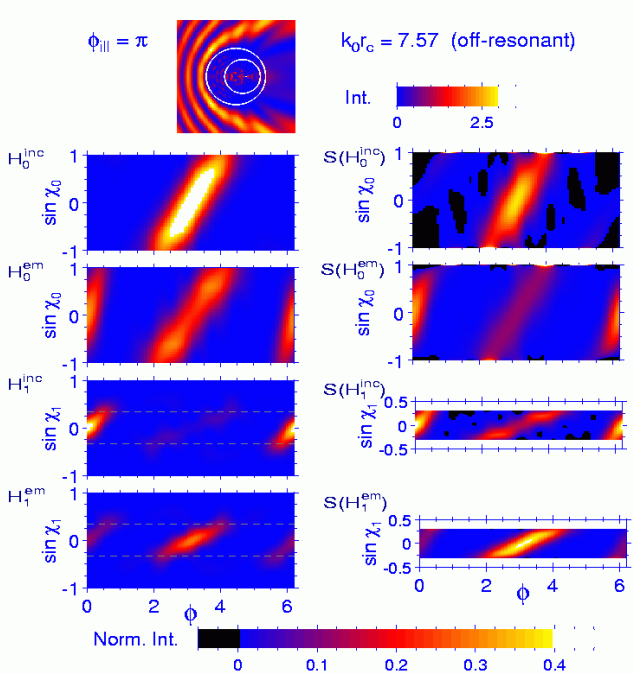}
  \caption{
Same as Fig. \ref{fig_diskoff}, but
for a dielectric annulus (refractive indices
  $n_0=1$ outside, $n_1=3$ in the annulus, $n_2=6$ in the inner disk,
radii $r_c=1,
  r_c'=0.6$, displaced by $\delta=0.22$).
  The system is illuminated by a plane wave at an off-resonant frequency
	(illumination direction $\phi_{\rm ill}=\pi$).
}
  \label{fig_husannbilloff}
\end{figure}

\begin{figure}[t]
\includegraphics[width=\columnwidth]{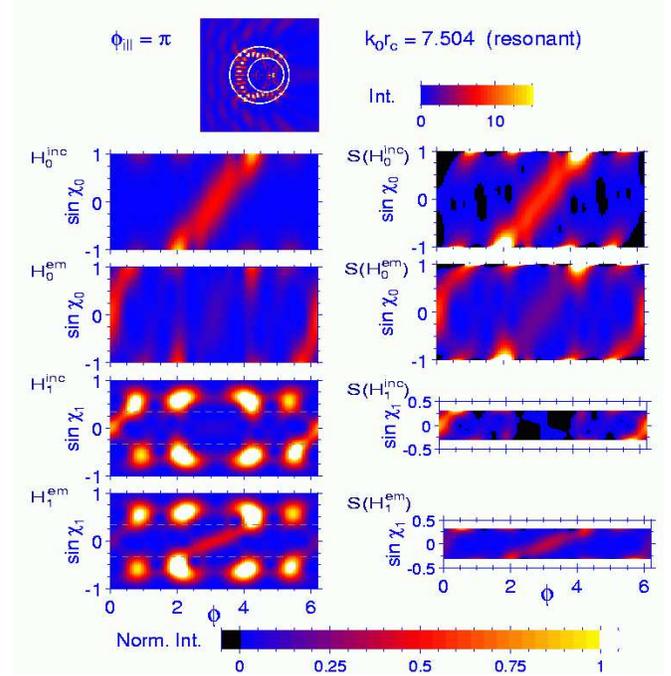}
  \caption{
Same as Fig. \ref{fig_husannbilloff}, but for illumination at a
  resonance frequency.
}
  \label{fig_annon}
\end{figure}

Finally, let us illustrate the usefulness of the Husimi functions
(\ref{hus}) also for a more complex system than the circular disk,
the annular system formed by regions of different refractive
indices that are confined by
two eccentric disks.
The ray optics in this system corresponds to  nonintegrable
dynamics in phase space, which allows for a multifaceted  set of
resonance wave functions \cite{annbillpre}.
Off resonance (Fig.\ \ref{fig_husannbilloff})
the scattering wave function enters the
dielectric system only barely, and the situation is similar to the
circular disk because the internal disk is not explored extensively.
At resonance the situation is very different.
Figure \ref{fig_annon} shows
a typical resonance wave function
in real space and its
Husimi representation in phase space.
The intensity of the resonance wave function is concentrated on
straight segments which can be identified as
a short
stable periodic trajectory of the corresponding
classical ray dynamics.
The Husimi functions display a strong intensity exactly
in the vicinity of this trajectory in classical phase space.

In conclusion, we introduced
four Husimi representations of the scattering wave
function at the interfaces of dielectric microresonators,
corresponding to the incident and emerging waves at both sides
of the interface. These Husimi functions  are easily computed
from the wave function and
have many desirable properties: They are related by the laws of Fresnel
and Snell in the ray-optics regime (i.e., short and off-resonant wavelength)
and generally
provide valuable detailed insight into the wave dynamics
in complex dielectric systems, most notably
even close to resonances where ray optics breaks down.

We gratefully  acknowledge helpful discussions with Jan Wiersig,
Arnd B{\"a}cker, and Christian Miniatura.


\begin{thebibliography}{99}
\bibitem{remark_pol}
For simplicity we only consider the case that the
electric field is polarized
perpendicular to the resonator plane.
\bibitem{remark_nonspherical}
All considerations
are easily extended
to non-spherical interfaces when one
replaces the polar angle
$\phi$ by the arc length $s$ along the interface.
\bibitem{remark_cutoff}
The unphysical singularity
${\cal F}^{-1}\to \infty$ for $\chi\to \pi/2$
is regularized by the corrections to the semiclassical derivation of Eq.\
(\ref{hus}). In our calculations we regularized ${\cal F}$
by setting a lower cutoff value ${\cal F}>(2/k)^{1/4}$, which is motivated by
these corrections.
\end{thebibliography}
\end{document}